\documentstyle[epsfig]{aipproc}

\newcommand{\xgo}{\mbox{$x_{\gamma}^{\rm obs}$}}

\begin{document}
\title{Introduction to high$-p_T$ inclusives}

\author{Matthew Wing$^{\dagger}$}
\address{
$^{\dagger}$McGill University,  Physics Department, \\ 3600 University Street, Montreal, \\
Canada, H3A 2T8 \\E-mail: wing@mail.desy.de}

\maketitle

\begin{abstract}
A selection of theoretical and experimental results are presented in the broader context 
of understanding QCD and its relation to photon physics. A phenomenological analysis of 
HERA data to constrain the gluon content of the proton and photon is discussed. 
Measurements from the Tevatron and fixed-target experiments are compared to theoretical 
predictions. Finally, the future of higher order pQCD calculations is addressed.
\end{abstract}

\section*{Introduction}

In these proceedings, advances in understanding QCD are discussed in experiments which 
directly complement those where measurements of the photon structure are being made. For a 
generalised accelerator, where the incoming particles, $I_1$ and $I_2$ resolve into 
partons, the cross section at leading order, $d\sigma_{I_1 I_2 \rightarrow cd}$, can 
be written as,
\begin{equation}
d\sigma_{I_1 I_2 \rightarrow cd} = \sum_{ab} \int_{x_{I_2}}
\int_{x_{I_1}} f_{I_2 \rightarrow b}(x_{I_2}, \mu_{I_2}^2) f_{I_1
\rightarrow a}(x_{I_1}, \mu_{I_1}^2) \mathcal{M}\mathit{_{ab \rightarrow cd}^2}, 
\label{eq1}
\end{equation}
where $f_{I \rightarrow b}(x_I, \mu_I^2)$, is the parton density function for a give 
momentum fraction, $x_I$ and factorisation scale, $\mu_I$ and 
$\mathcal{M}\mathit{_{ab \rightarrow cd}}$ is the $2 \rightarrow 2$ matrix element. 
This entails two or three unknowns; the perturbatively calculable matrix element 
and one or two structure functions. From equation~(\ref{eq1}) it can be seen that 
other experiments, such as the Tevatron, can provide complementary information on 
$\mathcal{M}\mathit{_{ab \rightarrow cd}}$ and therefore indirectly help measurements 
of the photon structure function.

\section*{HERA photoproduction data}

Improvements in the understanding of pQCD for jet photoproduction have led to the 
agreement bewteen independent calculations to within $5-10\%$~\cite{zeus95,nlo}. 
Considering the cross section as a function of pseudorapidity of one jet whilst 
restricting the other jet in the dijet system to a smaller region in pseudorapidity 
provides a good test of dijet production and the structure of the photon~\cite{zeus95}. 

In a recent paper~\cite{aur_et_al}, Aurenche et al. have considered using the HERA 
photoproduction data to constrain the gluon content of the photon and proton. The 
distribution in \xgo, the fraction of the photon's energy  
participating in the production of the two highest energy jets;
\begin{equation}
\xgo = \frac{\sum_{\rm jet1,2} E_T^{\rm jet} e^{-\eta^{\rm jet}}}{2yE_e},
\end{equation}
where $yE_e$ is the initial photon energy, was considered in different regions of 
pseudorapodity of the jet. Requiring two jets, $E_T^{\rm jet1,2}~>~12,~10$~GeV and the 
pseudorapidity, $0~<~\eta^{\rm jet}~<1$, the cross section shows a small dependence on 
the gluon distribution in the 
photon. However, when the jets are constrained to be more forward in pseudorapidity, 
$1~<~\eta^{\rm jet}~<2$, a significant sensitivity is seen at low$-\xgo$. A change of $30\%$ 
in the gluon density of the photon results in a $25\%$ change in the cross section at 
$\xgo~=~0.2$.

\begin{figure}[htb]
\unitlength=1mm
\begin{picture}(0,0)(100,100)
\put(142,53){\bf \Large{(a)}}
\put(230,56){\bf \Large{(b)}}
\end{picture}
\begin{center}
~\epsfig{file=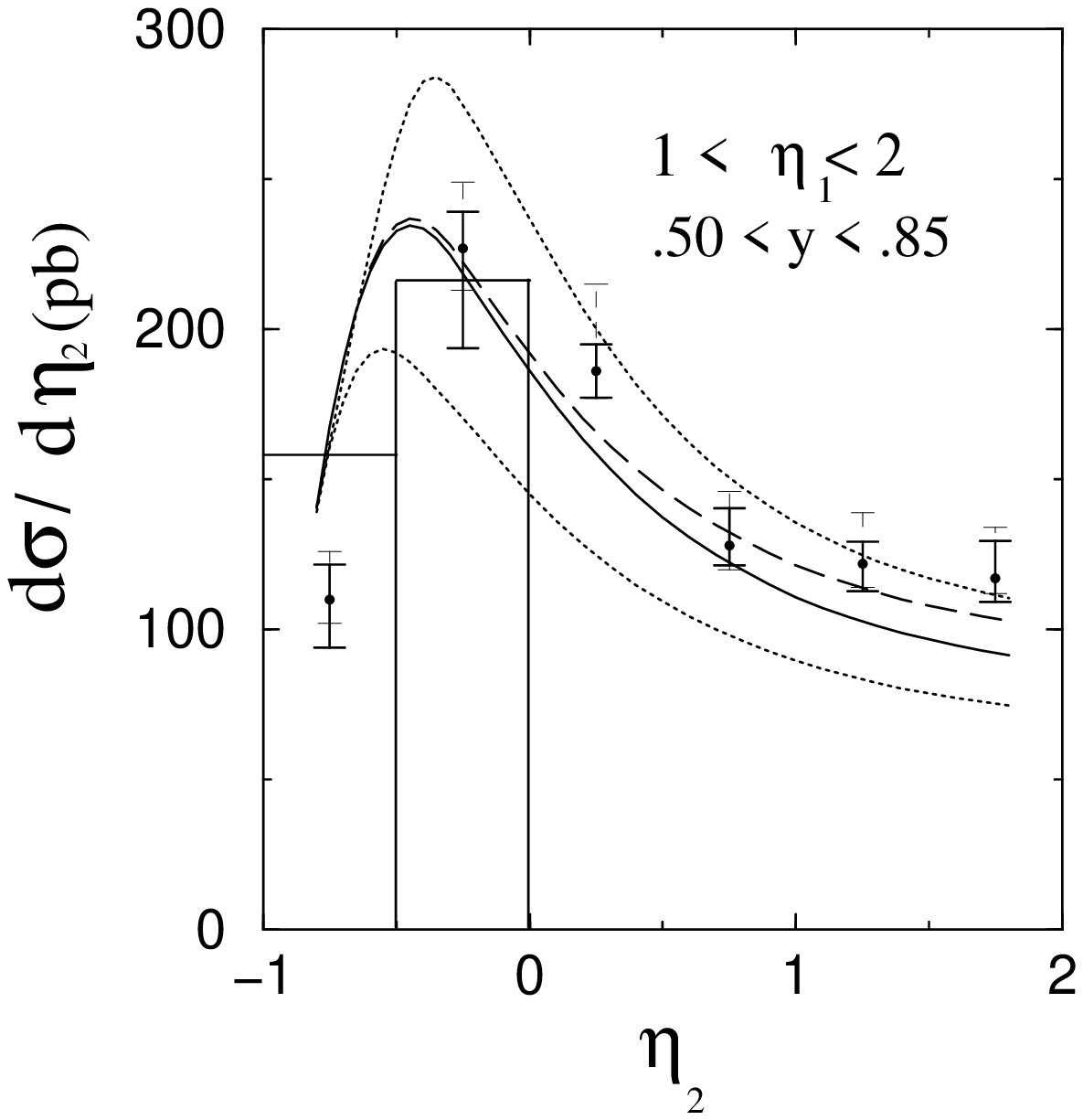,height=5.9cm}
~\epsfig{file=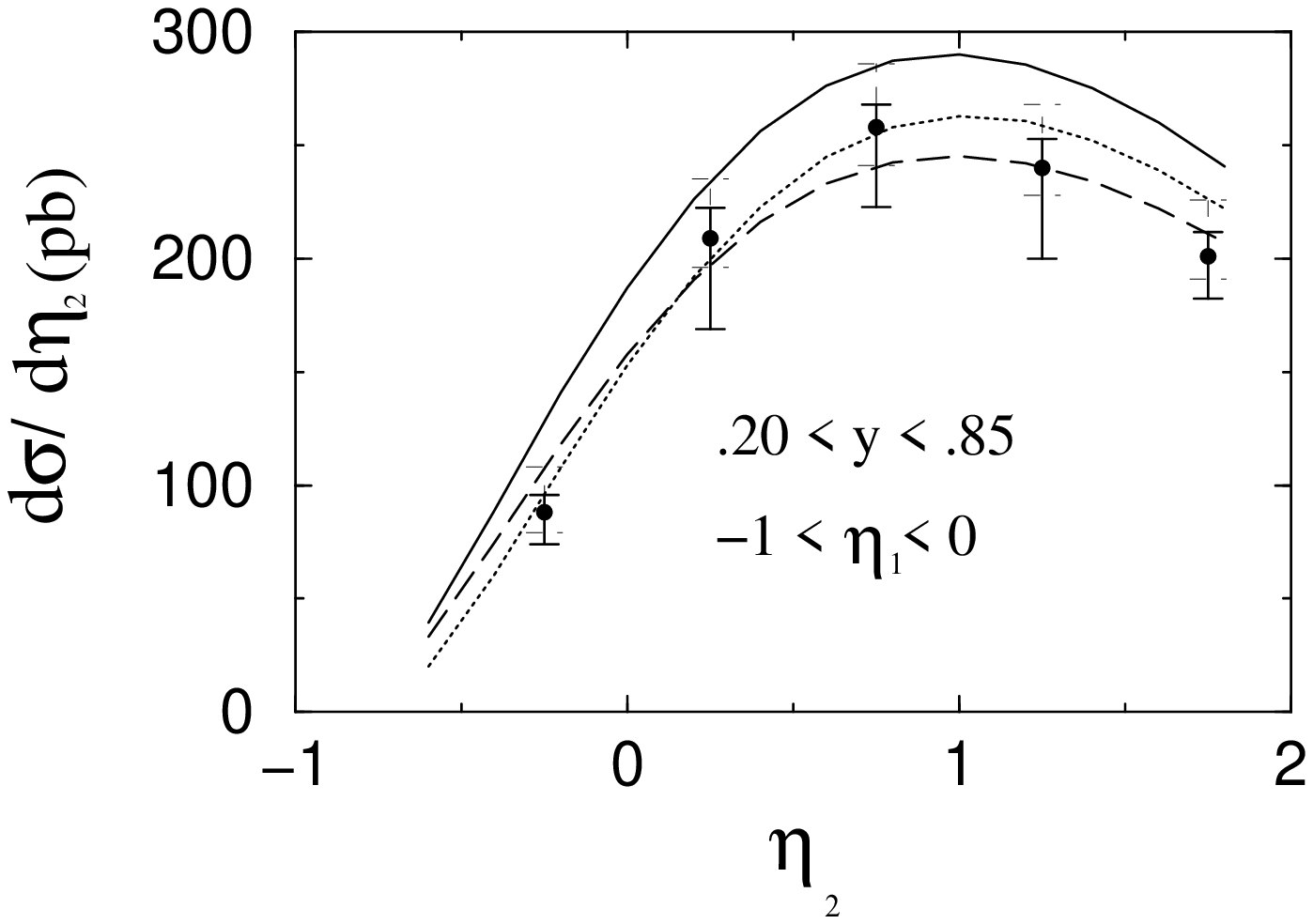,height=5.9cm}
\end{center}
\caption{The NLO cross section $d\sigma/d\eta_2$ compared to ZEUS data. In (a) the central 
prediction is the solid line and increasing the gluon content 
of the photon is the dashed line. In (b) the central prediction 
is the solid line and decreasing the gluon content of the proton is the dashed line. The 
dotted lines in (a) and (b) are changing the $y$ distribution (from~[3]).}
\label{eta_data}
\end{figure}

The NLO predictions are then compared to published data from the ZEUS 
collaboration~\cite{zeus95}, in which two jets, $E_T^{\rm jet1,2}~>~14,~11$~GeV, are 
required to be within $-1~<~\eta^{\rm jet}~<2$. The comparison is shown in 
Figure~\ref{eta_data}, where one jet is restricted to be in the forward region, 
$1~<~\eta^{\rm jet}~<2$ (Figure~\ref{eta_data}a) and the rear direction $-1~<~\eta^{\rm jet}~<0$ 
(Figure~\ref{eta_data}b).

In Figure~\ref{eta_data}a, increasing the gluon density in the photon by a value of $20\%$ 
improves the description of the data at large values of $\eta_2$, increasing the cross section 
by $\sim~10\%$. At negative values of $\eta_2$, the prediction remains the same and is larger 
than the data, however this region is subject to large hadronisation corrections. When the 
jet is in the rear direction as in Figure~\ref{eta_data}b, the cross section is less sensitive 
to the photon structure function and more so to the proton structure function. Decreasing 
the gluon density of the proton by $20\%$ in this region improves the description of the data. 
However, it should again be noted that the hadronisation corrections increase with decreasing 
$\eta_2$. 

Aurenche et al. conclude that the HERA data indicates a roughly $20\%$ increase in the 
gluon content of the photon and a $20\%$ decrease in the gluon content of the proton. The 
errors on the measurements are, however, as large as the change in cross section of these 
variations. Improved measurements are needed with larger statistics and are being 
investigated by both the H1 and ZEUS collaborations~\cite{h1,zeusnew}.

\section*{Tests of QCD}

Tests of QCD from other experiments, such as those at the Tevatron, complement those at HERA 
and LEP, with the 
Tevatron having the advantage of a better constrained incoming particle, the proton. 
Roughly, where HERA finishes; $E_T^{\rm jet}~\sim~50$~GeV and $M_{\rm jj}~\sim~200$~GeV, the 
Tevatron starts, providing overlap between the experiments at the different colliders. A 
selection of results of relevance to high$-E_T$ measurements at HERA and LEP will 
be discussed. In particular, high$-E_T$ jet production, jet substructure and prompt photon 
production will all be discussed later in the proceedings from at least one of the HERA 
or LEP experiments~\cite{bs,jt,cg}.

\subsection*{High$-E_T$ jet production}

\begin{figure}[htb]
\begin{center}
~\epsfig{file=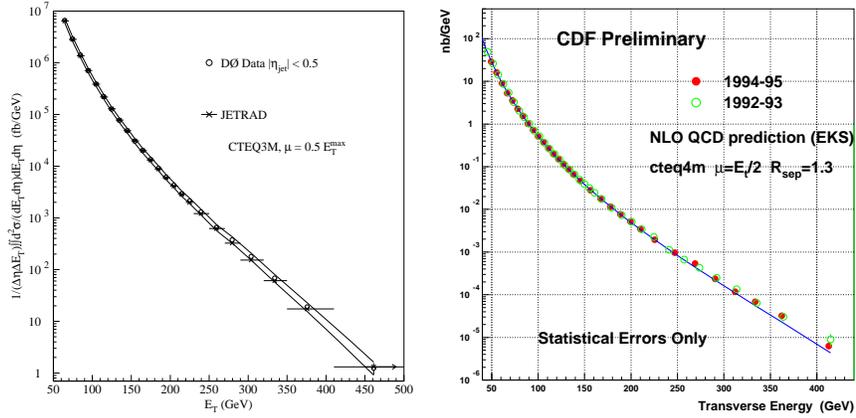,height=5.5cm}
~\epsfig{file=cdf_incl_jets.epsi,height=5.5cm}
\end{center}
\caption{Inclusive jet cross sections from the (a) D0 (from~[10]) and (b) 
CDF (from~[11]) collaborations at 
$\sqrt{s}~=~1800$~GeV. The measurements are compared to NLO predictions.}
\label{incl_cdf_d0}
\end{figure}

Inclusive jet measurements at the Tevatron extend to values of 
$E_T^{\rm jet}~\sim~450$~GeV~\cite{incl_jets}, 
falling by $\sim~7$ orders of magnitude as in Figure~\ref{incl_cdf_d0}. The NLO prediction 
describes the data from the D0 collaboration very well, with the CDF data lying above the 
prediction at high transverse energies, which could be a signal for new physics such 
as quark substructure. However, the proton structure function is less well constrained at 
these large scales and the question of whether the data can be accomodated in a change 
of the proton PDF arises.

In Figure~\ref{incl_3pdfs}, the data is compared to the same calculation with three different 
proton structure functions. The MRST PDF gives a generally poor description of 
both the CDF and D0 measurements at low and high transverse energies. The CTEQ4M proton PDF 
gives a good description of the D0 data but inadequately describes the CDF data at high 
transverse energies. The CTEQ4HJ PDF includes the CDF data from Run IA in its 
fit, so would be expected to describe the Run~IB data shown better than the other structure 
functions. This modified PDF describes well the D0 and the newer CDF data.

The comparison of data and theory clearly displays how understanding the parton density 
functions is essential in interpreting measurements in terms of new physics. Data to be 
and already analysed at HERA will allow the proton PDF to be further constrained at higher 
scales and hence more accurate predictions for the Tevatron measurements in Run II.

\begin{center}
~\epsfig{file=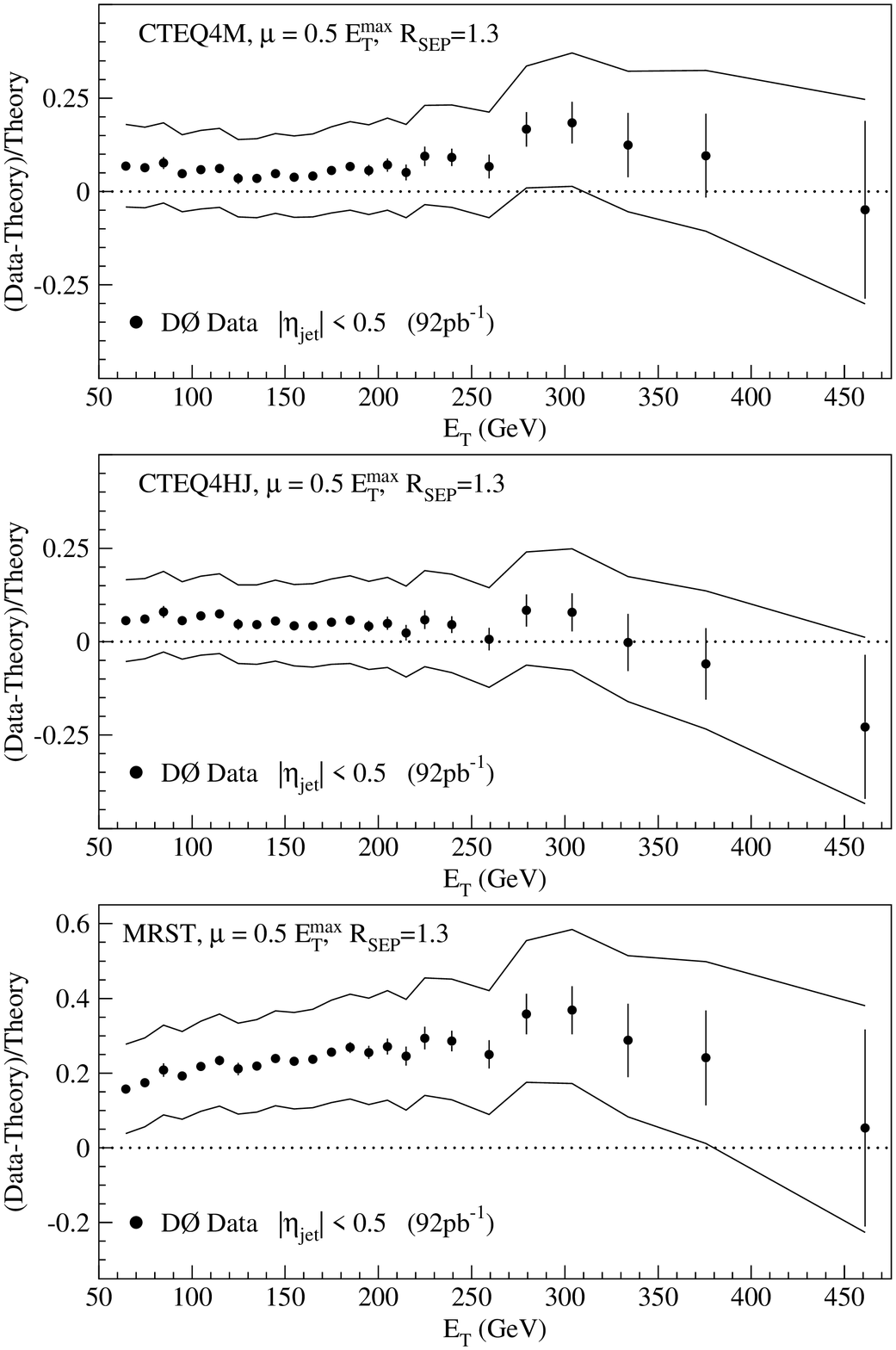,height=9cm}
~\epsfig{file=cdf_incl_jets_3pdfs.epsi,height=9.1cm,width=7.6cm}
\end{center}
\begin{figure}[htb]
\caption{Relative difference of data and theory for inclusive jet cross sections from 
D0~[10] and CDF (from~[11]) at $\sqrt{s}~=~1800$~GeV. The 
data is compared to three proton PDF's.}
\label{incl_3pdfs}
\end{figure}

\begin{figure}[htp]
\begin{center}
~\epsfig{file=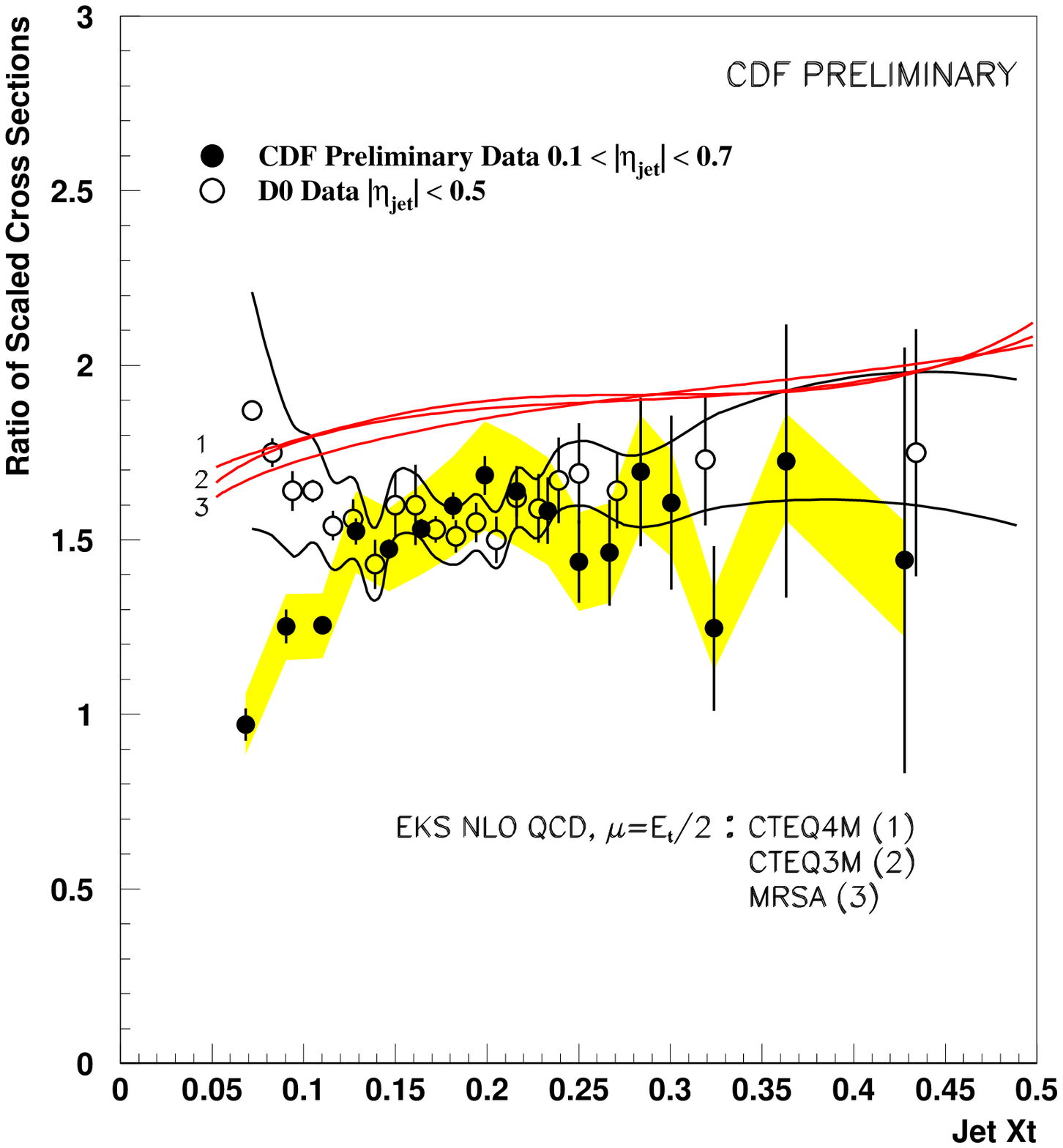,height=6.5cm,width=6.5cm}
~\epsfig{file=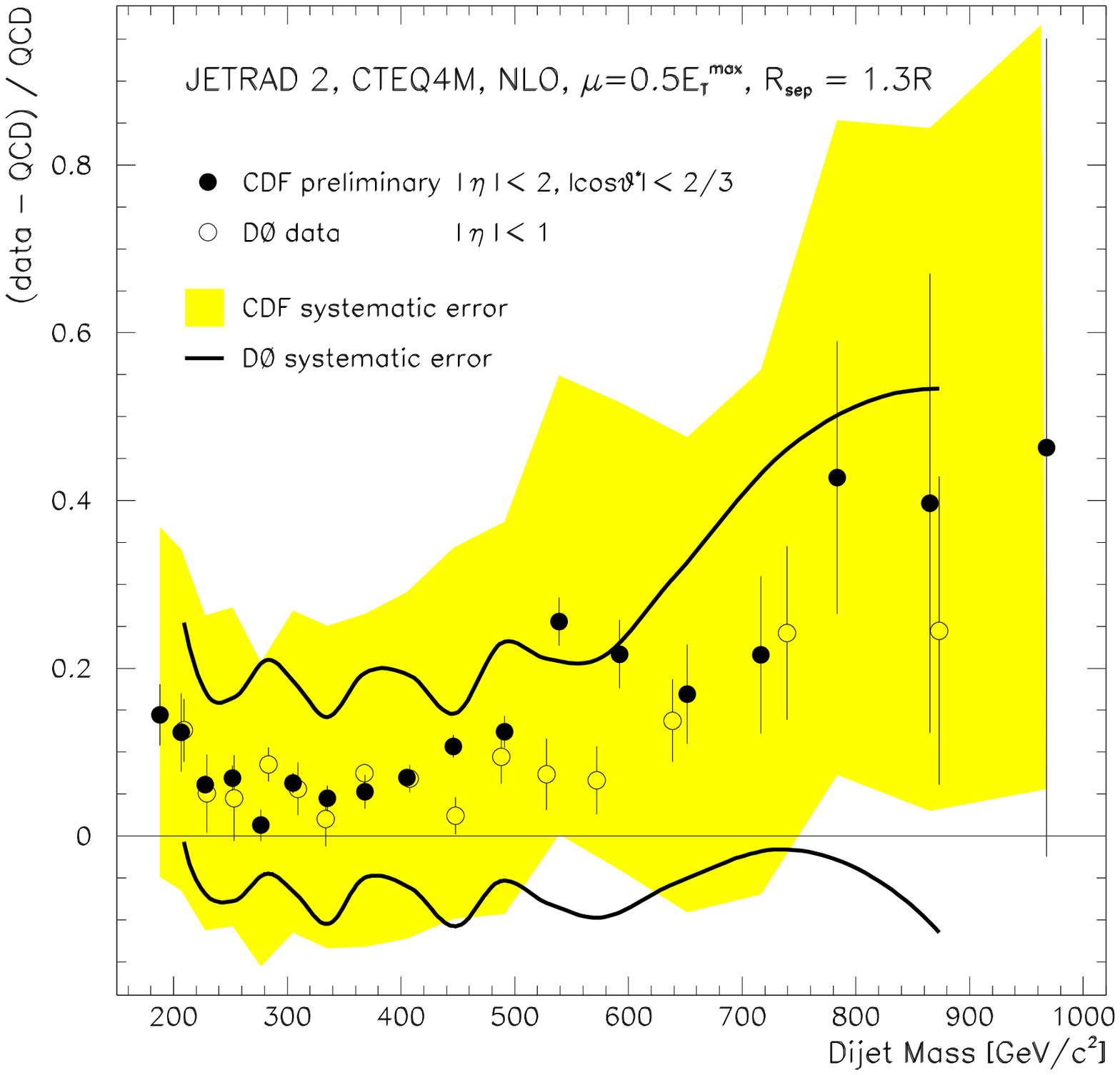,height=6.5cm}
\end{center}
\caption{(a) Ratio of scaled cross sections ($\sqrt{s}~=~630$~GeV$/\sqrt{s}~=~1800$~GeV) 
compared to an NLO calculation with three proton PDF's. (b) Relative difference of dijet 
invariant mass, measured by CDF and D0, compared with an NLO prediction (from~[11]).}
\label{s_ratio}
\end{figure}

The Tevatron also produced data at centre-of-mass energies, $\sqrt{s}~=~630$~GeV, lower than 
the nominal $\sqrt{s}~=~1800$~GeV. Comparing the cross sections at the two different energies 
provides a test of QCD whilst reducing systematic uncertainties on the measurement and the 
sensitivity to the choice of proton PDF. Figure~\ref{s_ratio}a shows the ratio of the inclusive 
jet cross sections as a funtion of $x_T~\equiv~\frac{2E_T}{\sqrt{s}}$ for both CDF and D0 data 
compared to NLO predictions. The reduction in the uncertainty in the choice of proton PDF can 
be seen in the difference of the prediction given by the three lines. At moderate and large 
$x_T$, the data agree well between the two collaborations, however at low $x_T$, the data sets 
diverge. At these lower values of $x_T$, the data suffer from possible problems of understanding 
the soft underlying event, which, both experiments correct for. The data also lie consistently 
below the prediction at the scale, $\mu$, chosen. Reasons for the discrepancy are unclear, with 
possible interpretations being the use of different renormalistion scales or $k_T$ effects, 
although neither of these are attractive explanations.

Consideration of the invariant mass of the dijet system, like the inclusive jet cross section, 
provides both a test of QCD and the opportunity to search for new physics. Figure~\ref{s_ratio}b 
shows the comparison of both experiments, CDF and D0, with an NLO prediction, where the two 
measurements are defined with jets in different angular regions. The measured data agree very well 
between experiments and with the theory. There is a tendency for the data 
to deviate from the prediction at high masses, but the systematic errors are too large to 
make any firm conclusions.

\subsection*{Jet substructure}

\begin{figure}[htb]
\begin{center}
~\epsfig{file=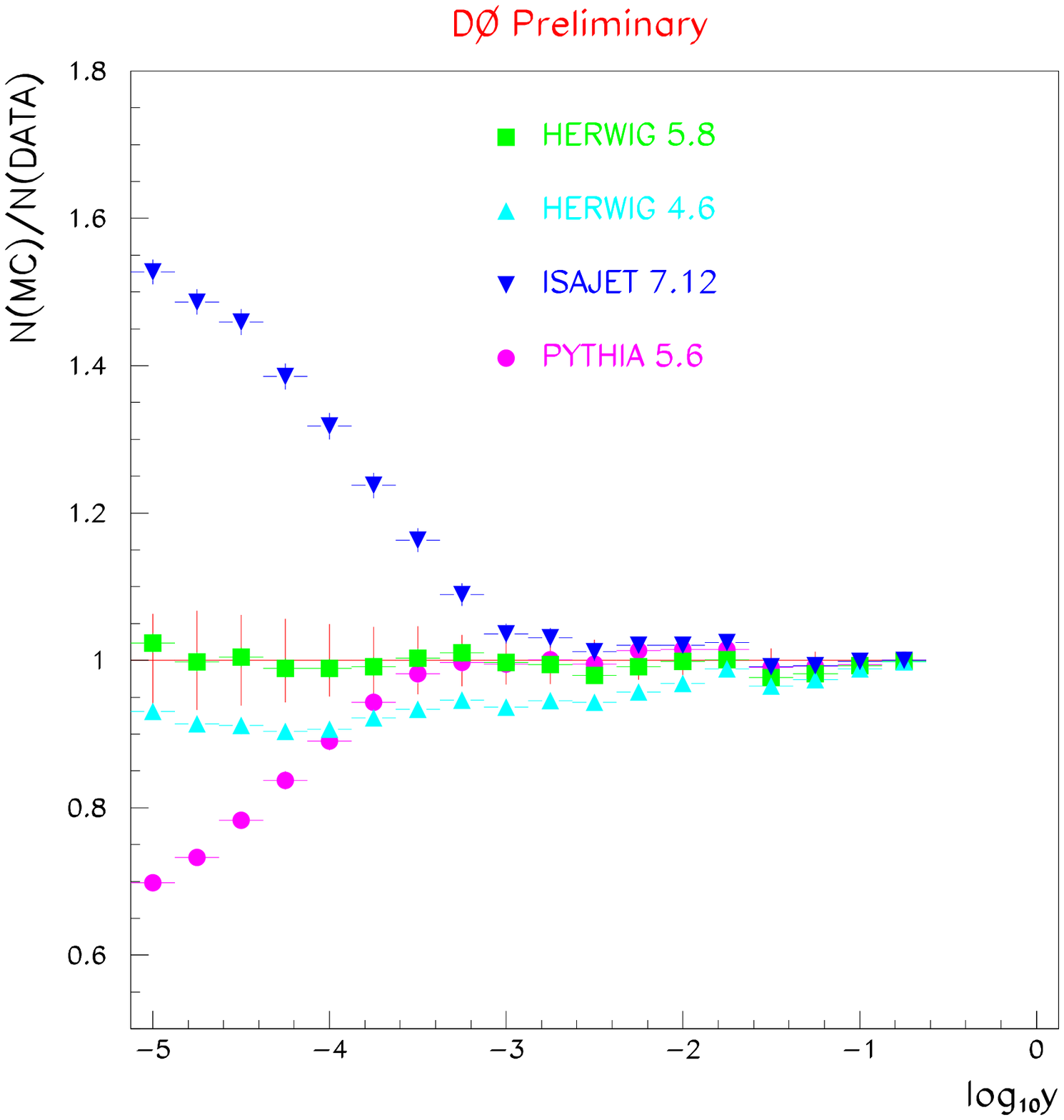,height=7.cm}
~\epsfig{file=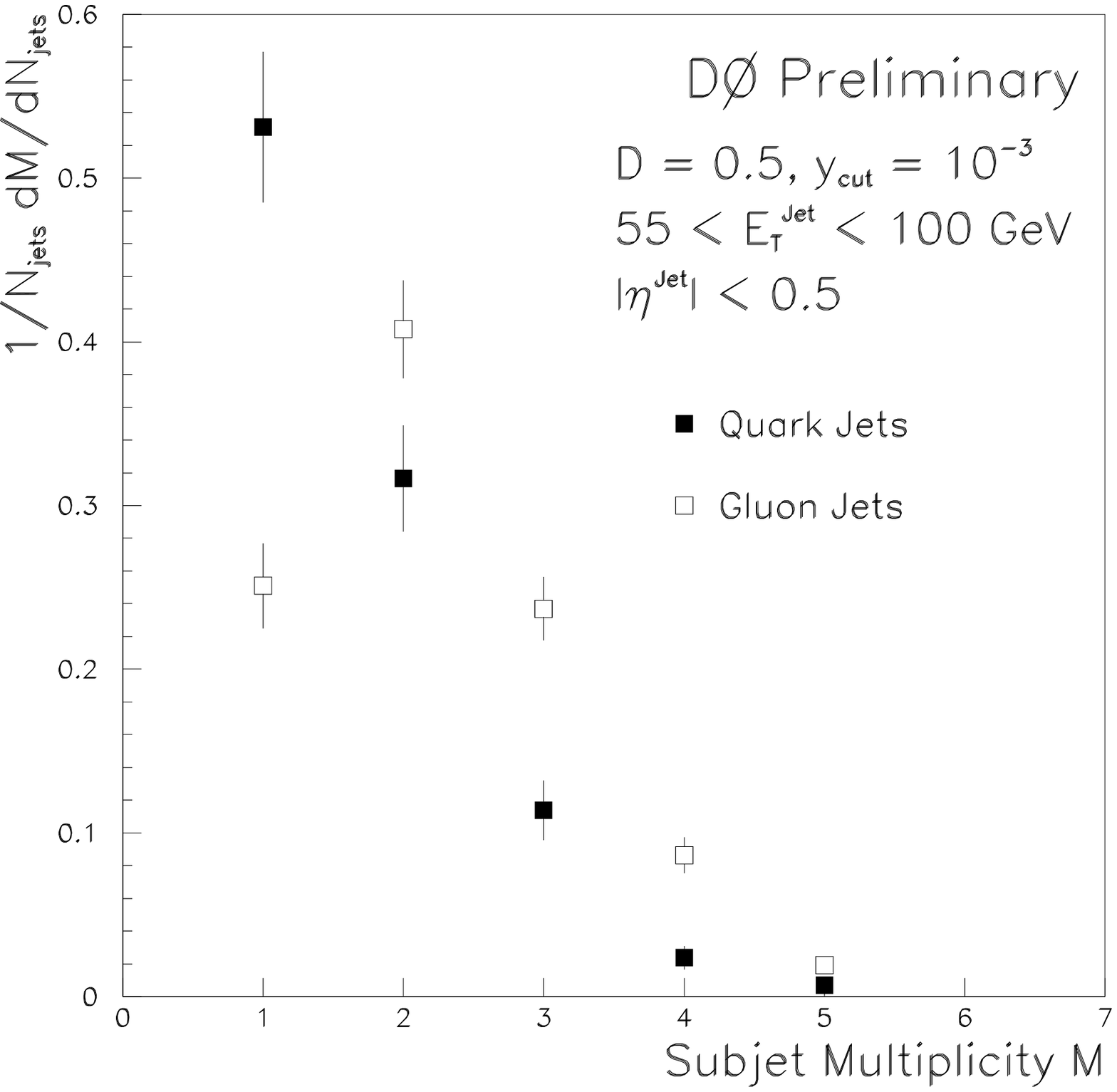,height=7.cm}
\end{center}
\caption{(a) Ratio of the multiplicity of subjets for data and MC for 
$E_T^{\rm jet}~>~250$~GeV (from~[12]). (b) Corrected subjet multiplicity in $q$ 
and $g$ jets extracted from D0 data (from~[13]).}
\label{subjets}
\end{figure}

Jet substructure has been studied at the Tevatron by considering both the jet shape and 
subjet multiplicity. Rerunning the $k_T$ algorithm on those particles assigned to jets and 
stopping the clustering when all values of $d_{ij}$ satisfy $d_{ij}~>~y_{\rm cut}E_T^2$ gives 
numbers of subjets as a function of $y_{\rm cut}$. Figure~\ref{subjets} show two different 
measurements of subjets using the $k_T$ algorithm~\cite{subjet,snihur}, where 
Figure~\ref{subjets}a shows a 
comparison of the number of subjets in data and different MC's and Figure~\ref{subjets}b 
shows how the subjet multiplicity can be used to differentiate between quark and gluon jets. 
The jets measured in Figure~\ref{subjets}a are above 250 GeV, which means that the scale 
being studied at the lowest $y_{\rm cut}$ is about 1 GeV. When one considers this range 
in scale, the description of the data by {\sc Herwig} is extremely good. Figure~\ref{subjets}b 
demonstrates a method for separating quark and gluon jets based on a statistical 
subtraction for events at $\sqrt{s}~=~630$~GeV and $\sqrt{s}~=~1800$~GeV. As expected, gluon 
jets show more activity than quark jets. A method for distinguishing quark and gluon jets 
has also been developed by the ZEUS collaboration~\cite{cg}. 

\subsection*{Prompt photon production}

\begin{figure}[htb]
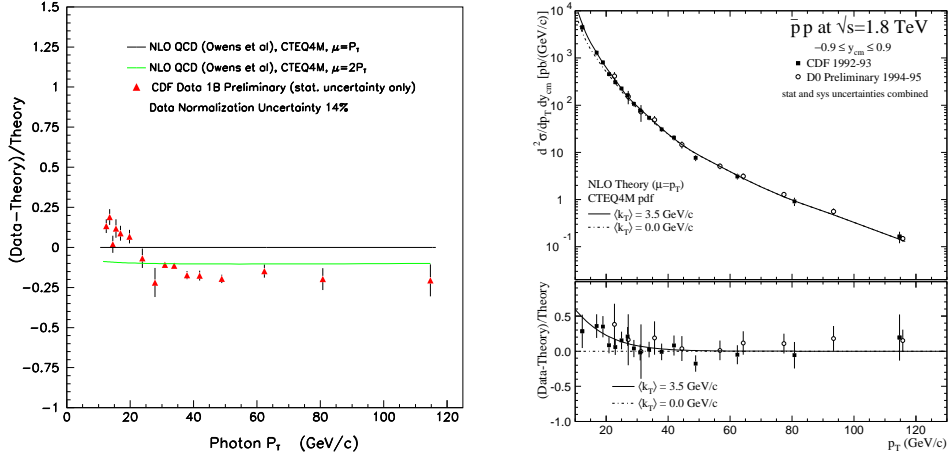

\begin{center}
~\epsfig{file=cdf_photon_owens_cteq4m.epsi,height=6.cm}
\hspace{0.5cm}
~\epsfig{file=figure2.epsi,height=6.cm}
\end{center}
\caption{(a) CDF prompt photon production data compared to NLO predictions 
(from~[11]). (b) CDF and D0 
data compared to NLO predictions with intrinsic $k_T$ and a value of $<k_T>~=~3.5$~GeV 
(from~[14]).}
\label{cdf_owens}
\end{figure}

Extensive measurements have been made and much theoretical work performed on prompt  
photon production at the Tevatron and fixed target experiments. Standard NLO perturbative 
calculations are unable to 
adequately describe the data at the Tevatron as shown in Figure~\ref{cdf_owens}a. Here, it 
can be seen that the data rise dramatically at low$-p_T$, the shape of which cannot 
be reproduced with a simple variation of the scale. The data can be described, however, 
by including an ``intrinsic $k_T$'' for the partons in the proton. Figure~\ref{cdf_owens}b 
shows the calculation rising at low$-p_T$ when a value of $<k_T>~=~3.5$~GeV is used~\cite{kt}. 
The use of intrinsic $k_T$ is somewhat unsatisfactory, differing between data sets and 
centre-of-mass energies.

\begin{figure}[htb]
\begin{center}
~\epsfig{file=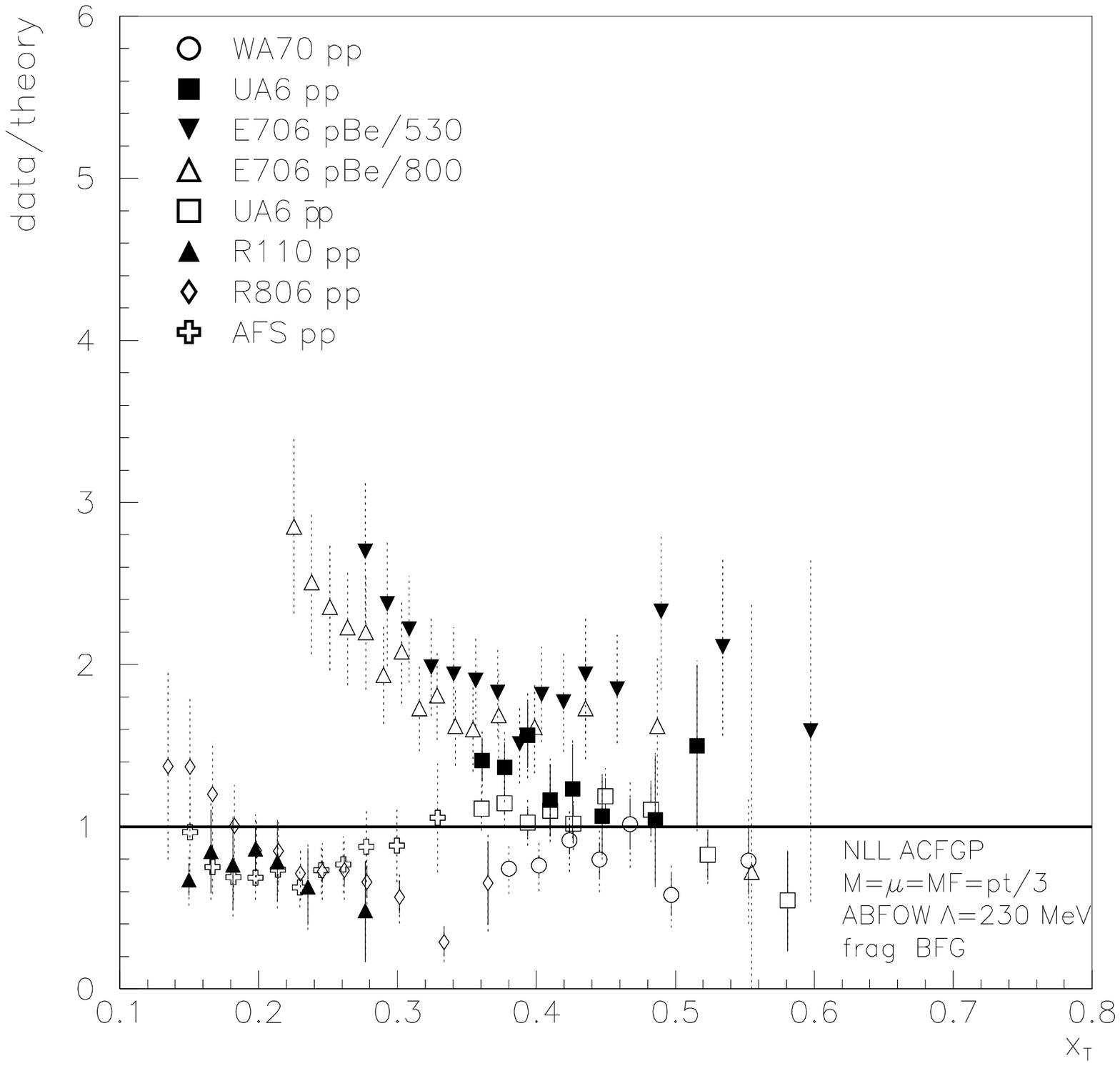,height=6.cm}
~\epsfig{file=e706comp.eps,height=5.5cm}
\end{center}
\caption{(a) Fixed target prompt photon data compared to NLO (from~[15]). 
(b) Data from E706 compared to NLO predictions which have resummed large 
logarithms in $x_T$ (from~[16]).}
\label{fixed_target}
\end{figure}

Prompt photon production has been extensively measured in fixed target experiments which 
have been compared with NLO predictions~\cite{aur_photon}. 
The NLO calculation from Aurenche et al~\cite{aur_photon} is able to describe all the fixed 
target data above a cut-off, $E_T~\geq~4-5$~GeV except that from E706. The ratio of data to 
theory is shown in Figure~\ref{fixed_target}a, where the E706 data is shown to dramatically 
rise at low $x_T~=~2p_T/\sqrt{s}$. 

Calculations resumming large logarithms in $x_T$~\cite{catani} are also unable to describe 
the E706 data as shown in Figure~\ref{fixed_target}b, although one would expect the 
resummation to improve the high$-p_T$ and not low$-p_T$ region. Figure~\ref{fixed_target}b 
also demonstrates how the resummation reduces the scale uncertainty in the calculation. 
Calculations in which $Q_T$, the nett transverse momentum of the final state $\gamma q$ pair, 
is resummed~\cite{laenen} improve the description of the E706 data at low $p_T$, although the 
NLO prediction is still too low. HERA data can provide useful informationon on intrinsic $k_T$, 
by filling the energy ``gap'' between the fixed target and Tevatron experiments~\cite{jt}.

\section*{Higher order calculations}

The last few years have seen large advances in producing higher order calculations for the 
production of two or three jets~\cite{kilgore,glover}. Calculations of NLO 3$-$jet 
hadroproduction are becoming available and NNLO 2$-$jet hadroproduction will be produced 
sometime in the future. The needs for a 3$-$jet NLO calculation are many~\cite{kilgore}. 
Measuring the 3$-$jet to 2$-$jet production ratio and hence $\alpha_s$ will be possible with 
a 3$-$jet NLO calculation. As well as testing QCD, the calculation will provide a better 
understanding of backgrounds to new physics processes. It is also an important step towards 
calculating NNLO 2$-$jet production.

\begin{figure}[hbp]
\begin{center}
\rotatebox{-90}{~\epsfig{file=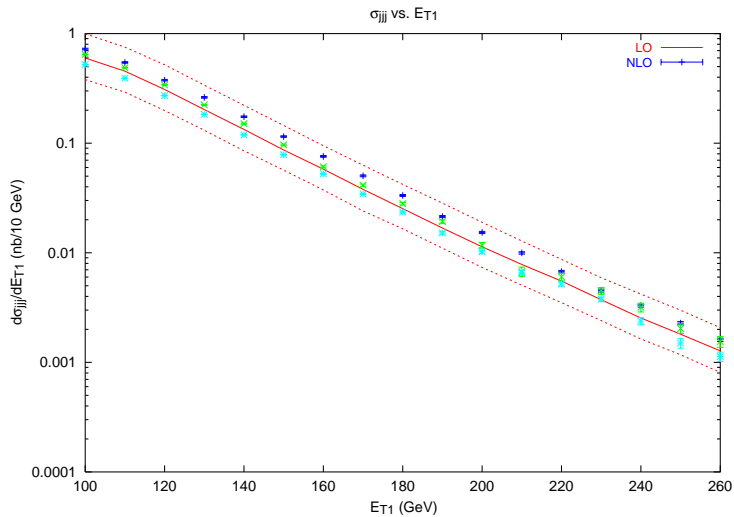,height=10cm}}
\end{center}
\caption{Predicted cross section for the highest-transverse-energy jet for 3$-$jet 
hadroproduction ($\sqrt{s}~=~1800$~GeV) where, $E_T^{\rm jet1}~>~100$~GeV, 
$E_T^{\rm jet2,3}~>~50$~GeV and $|\eta^{\rm jet}|~<~4$. The LO prediction with an estimtaion 
of the scale uncertainty is given by the lines and the NLO prediction with the corresponding 
estimtaion of the scale uncertainty is given by the points (from~[18]).}
\label{3-jet}
\end{figure}

Figure~\ref{3-jet} shows the predicted cross section for the highest-transverse-energy jet 
in 3$-$jet hadroproduction for both LO and NLO and their respective estimtaions of the scale 
uncertainty. The central NLO and LO predictions are of similar value, however, NLO 
shows a large reduction in the scale uncertainty over LO.

Progress on calculations of NNLO 2$-$jet hadroproduction has been good over the last two years 
and the interested reader is referred to recent talks on the subject~\cite{glover}. A NNLO 
calculation for the production of two jets is anticipated in a few years where a reliable 
estimate of the error on the cross section will be acheived.

\section*{Conclusions}

From the results presented, it can be seen that the theory of pQCD broadly describes the 
theory of the strong interaction. However, more detail both experimentally and theoretically 
is required to test QCD to great precision. This can also be acheived by considering 
the information from all experiments and what their results mean for each other.


\begin{references}

\bibitem{zeus95}
ZEUS Collab., J.Breitweg et al., \emph{Euro. Phys. J.} {\bf C11},~427~(1999).

\bibitem{nlo}
Frixione, S., {\it Nucl. Phys. B} {\bf 507}, 295 (1997);
Frixione, S. and Ridolfi G., {\it Nucl. Phys. B} {\bf 507}, 315 (1997);
Harris, B., Klasen M. and Vossebeld, J. {\tt hep-ph/9905348}; 
P\"{o}tter B., {\tt hep-ph/9911221}.

\bibitem{aur_et_al}
Aurenche, P., et al., {\tt hep-ph/0006011}.

\bibitem{h1}
H1 Collab., H1prelim-00-052, Submitted to ICHEP2000 , Osaka, Japan.\\
{\tt www-h1.desy.de/h1/www/publications/htmlsplit/H1prelim-00-052.long.html}

\bibitem{zeusnew}
ZEUS Collab., ICHEP$-$418, Submitted to ICHEP2000 , Osaka, Japan.\\
{\tt http://www-zeus.desy.de/$\sim$schlenst/conf/osaka\_paper/QCD/dijetpho.ps.gz};
ZEUS Collab., ${\rm EPS-540}$, Submitted to the EPS High Energy Physics 99 conference, 
Tampere, Finland. {\tt http://www-zeus.desy.de/eps99/eps99\_540.ps.gz}

\bibitem{bs}
Surow, B., (these proceedings)

\bibitem{jt}
Terron, J., (these proceedings)

\bibitem{cg}
Glasman, C. (these proceedings)

\bibitem{incl_jets}
CDF Collab., {\it Phys. Rev. Lett.} {\bf 77} 438.

\bibitem{d0_incl}
D0 Collab., {\it Phys. Rev. Lett.} {\bf 82} (1999) 2451.

\bibitem{blessed}
{\tt http://www-cdf.fnal.gov/physics/new/qcd/qcd99\_blessed\_plots.html}

\bibitem{subjet}
R. V. Astur, Proc. 10th Topical Workshop on Proton-Antiproton Collider Physics, 1995, 
eds. R. Raja and J. Yoh, p. 598.

\bibitem{snihur}
D0 Collab., {\tt hep-ex/9907059}, Submitted to the International Europhysics Conference 
on High Energy Physics, Tampere, Finland.

\bibitem{kt}
Apanasevich, L., et al., {\it Phys. Rev.} {\bf D59} (1999) 074007.

\bibitem{aur_photon}
Aurenche, P., et al., {\it Euro Phys. J.} {\bf C9} (1999) 107.

\bibitem{catani}
Catani, S., et al., {\it JHEP} {\bf 9903} (1999) 25.

\bibitem{laenen}
Laenen, E., Sterman, G. and Vogelsang, W., {\tt hep-ph/0006352} To appear in DIS2000 
conference proceedings.

\bibitem{kilgore}
Kilgore, W. and Giele, W., ``Hadronic three jet production  at NLO'', Talk presented at 
ICHEP2000, Osaka, Japan.

\bibitem{glover}
Glover, N., ``Jet cross sections in NLO QCD and beyond'', Talk presented at DIS2000, 
Liverpool, UK;
Bern, Z., Dixon, L. and Kosower, D. ``Recent progress in NNLO QCD calculation'', Talk presented 
at ICHEP2000, Osaka, Japan.

\end{references}
\end{document}